\documentclass[12pt]{article}
\pdfoutput=1
\usepackage[top=1in, bottom=1.25in, left=1.0in, right=1.0in]{geometry}
\usepackage[english]{babel}
\usepackage{atbegshi,cite}
\usepackage{amsmath,amssymb,amsbsy,amstext, amsthm, simplewick}
\usepackage{hyperref}
\usepackage{graphicx}
\usepackage{amsfonts}
\usepackage[small]{caption}
\usepackage{upgreek}
\usepackage[titletoc]{appendix}
\usepackage{setspace}

\newcommand{\newc}{\newcommand}
\newc{\fpi}{f_{\pi}}
\newc{\etap}{\eta^{\prime}}
\newc{\llll}{\langle\lambda\lambda\rangle}
\newc{\FFd}{F^a\tilde F^a}
\newc{\qbar}{{\overline q}}
\newc{\TR}{{\rm Tr}}
\newc{\Kahler}{K\"ahler }
\newc{\Rt}{{\mathbb R}^3}
\newc{\Rf}{{\mathbb R}^4}
\newc{\So}{{\mathbb S}^1}
\newc{\zt}{{\mathbb Z}_2}
\newc{\RtSo}{{\mathbb R}^3\times{\mathbb S}^1}
\newc{\scriminus}{{\cal I}^-}
\newc{\scriplus}{{\cal I}^+}
\newc{\mpl}{M_p}
\newc{\Ricci}{\mathcal{R}}

\begin{document}
\begin{titlepage}
\begin{flushright}
{\large 
ACFI-T16-33\\
}
\end{flushright}

\vskip 1.2cm

\begin{center}

{\large \bf Transplanckian Censorship and Global Cosmic Strings}

\vskip 1.4cm

{   Matthew J. Dolan$^{(a)}$, Patrick Draper$^{(b)}$, Jonathan Kozaczuk$^{(b)}$, and Hiren Patel$^{(b)}$}
\\
\vskip 1cm
{\it $^{(a)}$ARC Centre of Excellence for Particle Physics at the Terascale,\\
School of Physics, University of Melbourne, 3010, Australia  } \\
\vspace{0.3cm}
{\it $^{(b)}$Amherst Center for Fundamental Interactions, Department of Physics,\\ University of Massachusetts, Amherst, MA 01003
}\\
\vspace{0.3cm}
\vskip 4pt

\vskip 1.5cm

\begin{abstract}
Large field excursions are required in a number of axion models of inflation.  These models also possess global cosmic strings, around which the axion follows a path mirroring the inflationary trajectory. Cosmic strings are thus an interesting theoretical laboratory for the study of transplanckian field excursions. We describe connections between various effective field theory models of axion monodromy and study the classical spacetimes around their supercritical cosmic strings.  For small decay constants $f<M_p$ and large winding numbers $n>M_p/f$, the EFT is under control and the string cores undergo topological inflation, which may be either of exponential or power-law type. We show that the exterior spacetime is nonsingular and equivalent to a decompactifying cigar geometry, with the radion rolling in a potential generated by axion flux.  Signals are able to circumnavigate infinite straight strings in finite but exponentially long time, $t\sim e^{\Delta a/M_p}$.  For finite loops of supercritical string in asymptotically flat space, we argue that if topological inflation occurs, then topological censorship implies transplanckian censorship, or that external observers are forbidden from threading the loop and observing the full excursion of the axion.

\end{abstract}

\end{center}

\vskip 1.0 cm

\end{titlepage}
\setcounter{footnote}{0} \setcounter{page}{2}
\setcounter{section}{0} \setcounter{subsection}{0}
\setcounter{subsubsection}{0}
\setcounter{figure}{0}

\onehalfspacing



\section{Introduction}
\label{intro}

One of the fundamental observables associated with slow-roll inflation is the tensor-to-scalar ratio $r$, which can be constrained through measurements of the B-mode polarization of the cosmic microwave background~\cite{Ade:2015tva}. The Lyth bound connects large values of $r$ with large field excursions of the inflaton~\cite{Lyth:1996im}. With a number of new precision cosmological experiments currently active~\cite{Ade:2015tva,Ahmed:2014ixy} and in planning~\cite{Suzuki:2015zzg,Baumann:2008aq}, improving the theoretical understanding of large-field inflationary models has been a subject of much recent interest.

Scalars with approximate global shift symmetries are attractive candidates for the inflaton, as first pointed out in~\cite{naturalinflation}. In natural inflation, the inflaton is a compact modulus with scale of periodicity $f$. For the remainder of this paper we refer to such a field as an axion. An axion-inflaton traverses part of a circle of circumference $2\pi f$ in field space. However, natural inflation only proceeds if transplanckian axion excursions are allowed, implying $f>M_p$. Furthermore, at present, measurable values of $r$ also imply transplanckian field ranges for the inflaton. It is therefore of interest to construct controlled models of large-field axion inflation.

Extranatural inflation~\cite{extranatural} provided the first model realizing $f>M_p$ in a controlled effective field theory (with subplanckian energy densities and suppressed quantum gravity corrections). However, extranatural inflation resisted consistent completion within string theory~\cite{banksdinefoxgorbatov}, which led to the Weak Gravity Conjecture (WGC)~\cite{wgc}. For axion models, the WGC states that $f\cdot S_{inst} \lesssim M_p$, where $S_{inst}$ is an instanton action correcting the axion potential. Many effective field theories exhibiting long flat directions, including the extranatural inflation model, violate the WGC and are thought to live in the swampland~\cite{swampland,reece1,reece2}.

Recently there has been much work on stress-testing and sharpening the Weak Gravity Conjecture~\cite{reece1,reece2,Montero:2015ofa,Hebecker:2016dsw,Hebecker:2015zss,Hebecker:2015rya,Montero:2016tifx,Brown:2015lia,Brown:2015iha,Long:2016jvd,Bachlechner:2015qja,Bachlechner:2014gfa,Brown:2016nqt,Cottrell:2016bty}.
However, the connection between transplanckian flat directions in moduli space and consistency criteria like the WGC is not yet fully understood. The former are also associated with other curious ``censorship" properties in different settings. For example, in Ref.~\cite{nicolis}, Nicolis made the fascinating observation that even classical gravity might censor transplanckian field excursions from asymptotic observers (see more recently~\cite{Klaewer:2016kiy} on a similar topic). Ref.~\cite{nicolis} studied static, spherically symmetric scalar field sources, and found that classical GR places an absolute bound of order $M_p$ on the scalar variation outside the source. Inside the source, at least in the Newtonian approximation, transplanckian excursions appeared possible, but only in exponentially large experiments. In a wholly different context, Ref.~\cite{ArkaniHamed:2007js} showed that the low energy limit of certain string compactifications admits Euclidean wormholes when the moduli space contains a long geodesic. AdS/CFT shows that these saddle points do not contribute to the path integral, indicating that again transplanckian variations are censored.

Complementarily to extranatural inflation, axion monodromy models offer another controlled setting (from the EFT point of view) for the study of transplanckian field excursions. In these models, the axion's fundamental domain $f$ is smaller than $M_p$, but due to monodromies the axion can traverse this range $n$ times, leading to an effective excursion of order $n\cdot f >M_p$ during inflation. Inflationary models of this type were first proposed within string theory in~\cite{monodromy1,monodromy2}, and also have a wide variety of field theoretic realizations~\cite{knp,dante,kalopersorbo,dinedrapermonteux,maryland,fterm,Kappl:2014lra}. Some axion monodromy implementations are known to violate different versions of the WGC~\cite{reece1,reece2}.

A feature common to all axion models is the presence of global cosmic strings, and supercritical strings, or strings for which the product of the winding number $n$ and the symmetry breaking scale $f$ satisfies
$n\cdot f \gtrsim M_p$,
are a natural context in which to study large field variations. In monodromy inflation models, for example, cosmic strings of large winding number trace out an axion profile in {\it space} that is very similar to the inflationary trajectory explored by the axion in {\it time}. 
In this paper, we will use cosmic strings of large winding number and small $f$ as a controlled laboratory for the study transplanckian field excursions.  Our approach is ``bottom-up" and similar spirit to~\cite{nicolis}:  we will study the spacetimes associated with supercritical strings, and ask whether external observers have access to the large field excursions.

This paper is organized as follows. In Sec.~(\ref{sec:toymodels}), we review toy EFT models of axion monodromy inflation. We point out similarities and dualities relating the models and comment on features of their associated cosmic strings. 
In Sec.~(\ref{sec:cosmicstrings}) we study supercritical global cosmic string spacetimes in a representative 4D axion model and a 5D Wilson loop axion model. (Readers interested primarily in our results concerning cosmic string geometries can skip directly to this section.) In the controlled regime  $f<M_p$, we find analytical and numerical evidence that the string cores undergo topological inflation~\cite{Vilenkin:1994pv,Linde:1994hy} for $n\cdot f \gtrsim M_p$. The accelerated expansion is exponential in the 4D model and power-law in the Wilson loop case. We then obtain a candidate analytic solution for the spacetime exterior to an inflating string core. We show that it is equivalent to a decompactifying cigar geometry, with a radion potential sourced by axion flux, and find that it provides a good fit to the exterior spacetime in the numerical analysis of the 4D model. Finally, in Sec.~(\ref{sec:measure}), we ask whether causal trajectories can access the full axion field excursion around infinite cosmic strings and finite loops of string.  We conclude in Sec.~(\ref{sec:concl}) and comment on directions for future study. In Appendix A, we expand on a class of toy axion monodromy models based on chiral perturbation theory.

\newpage

\section{Toy Models of Axion Monodromy}
\label{sec:toymodels}
There are a number of toy models realizing axion monodromy in field theory. Examples include:
\begin{itemize}
\item The Abelian Higgs model in five dimensions
\item The four-form model of~\cite{kalopersorbo}
\item Large $N$ pure Yang-Mills, noted also in~\cite{kalopersorbo}
\item Multi-axion alignment models in various dimensions~\cite{knp,maryland}
\item Large $N$ supersymmetric Yang-Mills~\cite{dinedrapermonteux} and QCD(adj).
\end{itemize}

One differentiating feature between the models is whether or not the axion present at low energies is compact, meaning the theory respects a discrete gauge shift symmetry that acts on the axion alone, or noncompact, in which case there is still such a symmetry, but it acts also on other labels of the state. The abelian Higgs model on $R^4\times S^1$ is perhaps the simplest toy model that produces a noncompact axion at low energies. 

Without the charged scalar Higgs, $U(1)$ gauge theory on $R^4\times S^1$ gives rise to a compact scalar $a$ in the low energy 4D theory.  Forming the $U(1)$ holonomy 
\begin{align}
\Omega(x)=e^{ig\int_{S^1}A^5}\equiv e^{ia/f}\;,\;\;\; f^{-1}=2\pi g R\;,
\end{align}
 large gauge transformations imply $a\equiv a+1/gR$, where $g$ is the 4D gauge coupling. Furthermore, improper gauge transformations in the 5D theory result in a global shift symmetry in 4D, $a\rightarrow a+c$. These transformations are explicitly broken by the addition of charged matter $\Phi$, resulting in a 1-loop potential $V(a)$. However, $a$ remains compact, because large gauge transformations are preserved.  

When the $U(1)$ is spontaneously broken by $\langle\Phi\rangle=v$, the low energy axion develops a tree-level potential from the terms $g^2 v^2 (A^5)^2\rightarrow m^2a^2$. This ``decompactification" of the axion is due to monodromy~\cite{Silverstein:2013wua, McAllister:2014mpa, Furuuchi:2015foh}. In the Stueckelberg limit $\Phi\rightarrow ve^{i\alpha}$,  the field $\alpha$ can have winding number $k$ around the circle, and
the symmetry $a\rightarrow a+1/gR$ is restored when combined with a shift $k\rightarrow k-1$.

The axion in the four-form model of~\cite{kalopersorbo} is also noncompact. When embedded in five dimensions with a Wilson loop axion, the model is equivalent to the 5D Abelian Higgs model. The axion coupling to the four-form is realized in 5D as a Chern-Simons coupling:
\begin{align}
{\cal S}=\int~-F\wedge\star F-\frac{1}{2}F_4\wedge\star F_4+\mu A\wedge F_4+q\wedge(F_4-dC_3)\;,
\label{eq:Sfourform}
\end{align}
where $q$ is a 1-form Lagrange multiplier enforcing the Bianchi identity.
Integrating out $F_4$ and $C_3$ yields $\star F_4=q+\mu A$ and $q=d\varphi$, where $\varphi$ is a compact real scalar. The resulting action is
\begin{align}
{\cal S}=\int~-F\wedge\star F+\frac{1}{2}(d\varphi+\mu A)\wedge\star(d\varphi+\mu A)\;,
\end{align}
which is the Stueckelberg limit of the Abelian Higgs model.\footnote{Similar four-dimensional dualities appear in~\cite{Kaloper:2016fbr}.}

The large $N$ $\theta$ dependence of the pure Yang-Mills vacuum energy is thought to take the form:
\begin{align}
E_k(\theta)\simeq\Lambda^4(\theta+2\pi k)^2+{\cal O}(1/N)\;,
\label{EkYM}
\end{align}
with $k$ labeling different branches~\cite{Witten:1980sp,Witten:1998uka}. 
In the large $N$ expansion, axions are again noncompact at energies far below $\Lambda$. The monodromy responsible for the decompactification is related to a change in topological charge density, as is evident from the $\theta$-derivative of Eq.~(\ref{EkYM}): as $\theta\rightarrow\theta+2\pi$, $F\tilde F\rightarrow F\tilde F+c\cdot \Lambda^4$. In the ultraviolet, the axion can again be embedded as a $U(1)$ holonomy, with mixed Chern-Simons coupling given by the replacement $\mu A\wedge F_4\rightarrow A\wedge F^a\wedge F^a$ in Eq.~(\ref{eq:Sfourform}). Indeed, as noted in~\cite{Gabadadze:2002ff}, the four-form theory has an interpretation as a 1PI effective action for the topological charge density. 

In the noncompact axion monodromy models, there are global cosmic strings associated with the discrete gauge symmetries. These strings are termination lines for domain walls: as the axion winds around the string, it has to jump across a potential barrier set by an ultraviolet scale. For example, in large $N$ YM, there is a nonperturbative potential of order $\Lambda^4$ separating the branches at fixed $\theta$. The string must pass through this region in order to return to a gauge-equivalent state, so the configuration acquires a domain wall. Strings of winding number $n$ acquire $n$ domain walls. 

The domain walls are not part of the inflationary trajectory, and it is convenient to take a simplifying limit in which they are eliminated. This can be achieved by setting the inflaton potential to zero (e.g., $\Lambda\rightarrow 0$ in YM). 
In noncompact axion models, this limit also corresponds to eliminating the monodromy from the IR theory. However, cosmic strings of large winding number still realize transplanckian excursions of the axion, which can be identified with the limiting trajectory of the axion during inflation as the potential is turned off.

A simple class of compact axion monodromy models was described in~\cite{knp,maryland}. The Chern-Simons models of~\cite{maryland} take the 5D form
\begin{align}
{\cal L}\sim (k A+B)\wedge F^a\wedge F^a+B\wedge G^a\wedge G^a
\end{align}
where $A$ and $B$ are U(1) gauge fields and $k$ is a large number. The low energy potential for the two holonomies ($\alpha$ and $\beta$, both $2\pi$ periodic) is taken to have the form 
\begin{align}
V\sim V_0^F\cos(k\alpha+\beta)+V_0^G\cos(\alpha).
\end{align}
 The simplest case to analyze occurs when the scale of $F$ is much greater than the scale of $G$, in which case $\alpha$ can be integrated out. 
The low energy theory contains $\beta$, which is still periodic, but with $2\pi k\gg2\pi$ periodicity. The residual inflationary potential is $V_0^G\cos(\beta/k)$. These models are clearly very similar to YM at large $N$, but they can be studied at small $N$ as well, at the expense of an additional large parameter. At low energies, $k$ appears as a large anomaly coefficient coupling the Wilson loop axion $\alpha$ to $F\tilde F$, and it is important that there is another anomaly of comparatively small coefficient.

The supersymmetric QCD example studied in~\cite{dinedrapermonteux} shares elements of the large $N$ YM model and of the multi-axion models, although it is a compact model with only one axion. Pure SYM has a discrete $\mathbb{Z}_{2N}$ $R$-symmetry spontaneously broken to $\mathbb{Z}_{2}$ by the gluino condensate, 
\begin{align}
\langle \lambda\lambda\rangle=32\pi^2\Lambda^3 e^{\frac{2\pi i k +\theta}{N}},
\end{align}
with $k$ labeling the resulting $N$ vacua. The $k$-vacua are smoothly traversed under $\theta\rightarrow\theta+2\pi$, giving rise to monodromy.  
When $\theta$ is promoted to an axion or an $\eta^\prime$, the potential is flat 
and the axion has periodicity $2\pi N$. 
This can also be understood as a consequence of having two anomalies with hierarchically different coefficients ($U(1)_{PQ}$, with anomaly coefficient of order $1$, and $U(1)_R$, with anomaly coefficient of order $N$.) This model is therefore similar to a limit of the previous model in which $\alpha$ becomes strongly coupled, analogous to the degrees of freedom in ${\rm arg~}\lambda\lambda$, and $k\rightarrow N$.

A small breaking of the discrete $R$-symmetry, for example through a soft gluino mass $m_\lambda$, generates a potential of the form 
\begin{align}
V=m_\lambda \Lambda^3\cos\left(\frac{a}{Nf}\right)\;.
\end{align}
(The saxion direction must be stabilized by a soft SUSY-breaking mass, which preserves the $R$ symmetry.) Generalizations of the SQCD model to nonsupersymmetric models with multiple adjoint fields are discussed in Appendix~\ref{appx}.

In compact axion monodromy models, there is a discrete gauge symmetry that acts only on the axion of the low-energy theory, and in contrast to the noncompact axion models, monodromy is retained as the inflaton potential is turned off. For example, in the SYM model, the discrete gauge symmetry is $a\rightarrow a+2\pi Nf$, and the axion potential is flat in the $m_\lambda\rightarrow 0$ limit. Therefore, in this limit there is a cosmic string associated with $a$ that follows the inflationary trajectory and has no attached domain walls. From the point of view of the infrared theory, the winding number of the axion is $1$, and the symmetry-breaking scale is $f_{eff}=Nf>M_p$. From the point of view of the microscopic theory, the winding number of the axion is $N$, the winding number of ${\rm arg~}\lambda\lambda$ is $1$, and the symmetry-breaking scale is $f<M_p$. For $\Lambda\ll f$, the energy densities in the string are primarily associated with the axion, and the presence of a varying ${\rm arg~}\lambda\lambda$ is a perturbation.

To summarize, each of these simple, closely-related field theory settings for axion monodromy inflation contains global cosmic strings around which the axion follows a trajectory similar to its trajectory during inflation. These are \emph{supercritical} strings, or strings for which the product of the winding number and the decay constant satisfies
\begin{align}
n \cdot f \gtrsim M_p\;,
\end{align} 
and they furnish a natural, controlled laboratory for the study of transplanckian field excursions.
 In the limit that the inflationary potential vanishes, the cosmic strings do not have attached domain walls. In the next sections, we discuss the spacetime properties of supercritical strings.

\section{Supercritical Cosmic Strings}
\label{sec:cosmicstrings}
Like a string in flat space, the gravitating string can be described in terms of a core region, where a noncompact modulus associated with the axion is excited away from its vacuum value, and an exterior region, where this radial mode is frozen and the local stress-energy is due to the angular gradient of the winding axion. 
A number of works have studied the spacetimes in and around global cosmic strings~\cite{Harari:1988wa,Cohen:1988sg,Gregory:1988xc,Vilenkin:1994pv,Linde:1994hy,Gregory:1996dd,Cho:1998xy}. The first complete solution to the Einstein equations in the exterior region was given by Cohen and Kaplan in~\cite{Cohen:1988sg}. The CK metric is static and exhibits an angular deficit increasing as a function of radius, ending at a singularity. For strings with $f<M_p$, the singularity is exponentially far from the core, but for strings with $f>M_p$, it moves inside the core. In~\cite{Gregory:1996dd}, Gregory obtained a nonsingular solution for $f<M_p$ strings arising in $\Phi^4$ theory by relaxing the staticity requirement, allowing inflation in the direction parallel to the string but holding the proper radius of the core region fixed.  Earlier, Vilenkin and Linde argued that even the fixed radius condition must be relaxed for global defects with $f>M_p$, allowing the cores to undergo topological inflation both axially and radially~\cite{Vilenkin:1994pv,Linde:1994hy}. This was suggested to be a consistent limit for the subcritical nonsingular string solutions in $\Phi^4$ theory~\cite{Gregory:1996dd,Gregory:2002tp}, and was confirmed in a numerical study by Cho~\cite{Cho:1998xy}.

We will build on these analyses in several directions:
\begin{enumerate}
\item  By considering  $f\ll M_p$ and $n\cdot f> M_p$, 4D EFTs like $\Phi^4$ theory are under better theoretical control, in the same way that the toy models of axion monodromy inflation discussed in the previous section offer greater  control than 4D models of natural inflation.
The numerical study in Ref.~\cite{Cho:1998xy} investigated low values of $n$ and suggested that the critical value of $f$ for which topological inflation occurs decreases as the winding increases. We will understand this analytically in $\Phi^4$  theory, showing that the relevant parameter is $f_{eff}=n\cdot f$. We numerically solve the Einstein-scalar field equations out to large values of $n$, and verify that exponential expansion may still proceed for $f\ll M_p$ once $n$ is sufficiently large. 
\item  The spacetime around vortices associated with 5D Wilson loop axions (which we take to be representative of a larger class of compact moduli associated with higher dimensional gauge fields) have received less attention. Typical potentials for the noncompact modulus have a qualitatively different form from $\Phi^4$ theory, and it is less clear what happens inside the core of supercritical strings. We will show that they can also undergo topological inflation, but the expansion is power law in time rather than exponential. 
\item To our knowledge, the spacetime exterior to topologically inflating global strings has not been determined. We find a candidate metric, which has a dual interpretation as a circle compactification with time-dependent radion, and we show that it matches onto the numerical simulation of the $\Phi^4$ string spacetime.\footnote{We note that there has also been recent progress in the study of spacetimes associated with local supercritical strings~\cite{Niedermann:2014yka}, and the geometry found in~\cite{Niedermann:2014yka} shares a number of common features with what we will find for the global string.}
\end{enumerate}

\subsection{Infinite String Interior}
\label{sec:core}
\subsubsection{4D Axions}
$\Phi^4$ theory provides a toy model for the conventional 4D cosmic strings discussed in Sec.~(\ref{sec:toymodels}). To fix notation, we take
\begin{align}
V(\phi)=\frac{\lambda}{4} (|\Phi|^2-f^2)^2\;,~~~\Phi=\phi e^{ia/f}\;,~~~a=n\theta f \;,
\label{phifour}
\end{align}
where $f$ is the scale of symmetry breaking and $n$ is the winding number. We also define
\begin{align}
\epsilon\equiv \frac{f^2}{M_p^2}
\end{align}
where $M_p$ is the reduced Planck mass. We are interested in the parameter region $f\ll M_p\ll nf$.

Away from the core of the string, the radial mode $\phi$ can be integrated out, $\phi\rightarrow f$. In the string core, $\phi\rightarrow 0$ as $r^n$.  If the core of the string inflates, the field $\phi$ near the boundary of the core no longer adequately balances gradient and potential energy, and $\phi$ begins to collapse toward its vacuum. Let us assume that the core boundary collapses in this way and derive consistency conditions for the presence of topological inflation. Let us further assume the evolution of the scalar field near the boundary can be estimated from the flat space equation of motion, linearized around the vacuum field. We will see what this assumption implies below.

In the linearized flat-space approximation, the scalar field on the boundary of the core grows exponentially as
\begin{align}
\phi(t)=\phi(t_0)e^{\sqrt{V^{\prime\prime}(f)}t}\;
\label{eq:phit}
\end{align}
If we define the boundary by a radius $r_k(t)$ satisfying $\phi(t,r_k(t))=kf$, where $k$ is a small parameter we choose by hand, then using $\phi(t_0,r)\simeq c r^n$ at a fixed time $t_0$, we obtain
\begin{align}
r_k(t)\sim e^{-\frac{\sqrt{V^{\prime\prime}(f)}}{n}t}\;.
\end{align}
 Meanwhile, far inside the core the potential density is approximately constant. The Hubble radius is
\begin{align}
H^{-1}=\frac{\sqrt{3}M_p}{\sqrt{V(0)}}\;
\label{hubble}
\end{align}
Refs.~\cite{Vilenkin:1994pv,Linde:1994hy} noted that inflation may occur in the core if the Hubble radius is smaller than the core size. In flat space, the core radius may be estimated by equating the potential at the top of the hill to the axion gradient energy,
\begin{align}
r_{core}\simeq \frac{nf}{\sqrt{V(0)}}\;,
\end{align}
which, comparing to Eq.~(\ref{hubble}), suggests that topological inflation occurs if 
\begin{align}
n\sqrt{\epsilon} \gtrsim 1\;.
\label{topoinfl}
\end{align}
Another estimate is obtained by comparing the rate of collapse of the coordinate boundary of the core to the expansion rate in the interior, and leads to the same result. The proper radius to the core boundary $r_k$ evolves with time as
\begin{align}
d(r_k)\simeq  e^{\left(H-\frac{\sqrt{V^{\prime\prime}(f)}}{n}\right)t}\;.
\label{eq:propd}
\end{align}
A positive exponent indicates inflation, or accelerating growth of the interior area, and is equivalent to the condition in Eq.~(\ref{topoinfl}). Since $H$ does not depend on $n$, for any values of parameters, there is an $n$ such that exponential growth occurs. In particular, even for significantly subplanckian $f$, topological inflation occurs so long as $nf>M_p$.

Note that in $\Phi^4$ theory, we could equally well have linearized the flat space equation of motion around $\phi=0$. In that case we can rewrite the proper radius as
\begin{align}
d(r_k)\simeq  e^{\left(1-\frac{\sqrt{V^{\prime\prime}(0)}}{n M_p\sqrt{V(0)}}\right)Ht}\;,
\label{eq:propd2}
\end{align}
where the slow-roll parameter $\eta$ appears in the exponent as $\eta/n$. Subplanckian $f$ corresponds precisely to a {\emph{large}} value for $\eta$, justifying our flat-space approximation in Eq.~(\ref{eq:phit}).\footnote{The analysis here is similar to one given by Vilenkin in~\cite{Vilenkin:1994pv} for the case of the supercritical domain wall with $\epsilon>1$. In that case, it is appropriate to follow scalar field evolution near the boundary of the wall using the slow roll equation of motion, $3H\dot{\phi}=-V^\prime(\phi)\approx \lambda f^2\phi$. We are instead interested in the regime $\epsilon\ll1$, where we can drop the Hubble friction term and solve the flat space equation of motion. The string also introduces new dependence on the parameter $n$.}  The additional $1/n$ suppression allows topological inflation to proceed, justifying our initial assumption that the string profile would collapse. Subplanckian $f$ also keeps effective field theory under control, leading to small curvatures everywhere.

A negative exponent in Eq.~(\ref{eq:propd}) or~(\ref{eq:propd2}), on the other hand, does not imply exponential contraction of the proper distance. When the core collapse is able to keep pace with the  growth of metric components in the core, the string is supported by gradient energy as it is in flat space. In this case the proper distance should saturate to a constant, smoothly connecting onto nonsingular metrics for subcritical strings~\cite{Gregory:1996dd}.

\begin{figure}[t!]
\begin{center}
\includegraphics[width=0.45\linewidth]{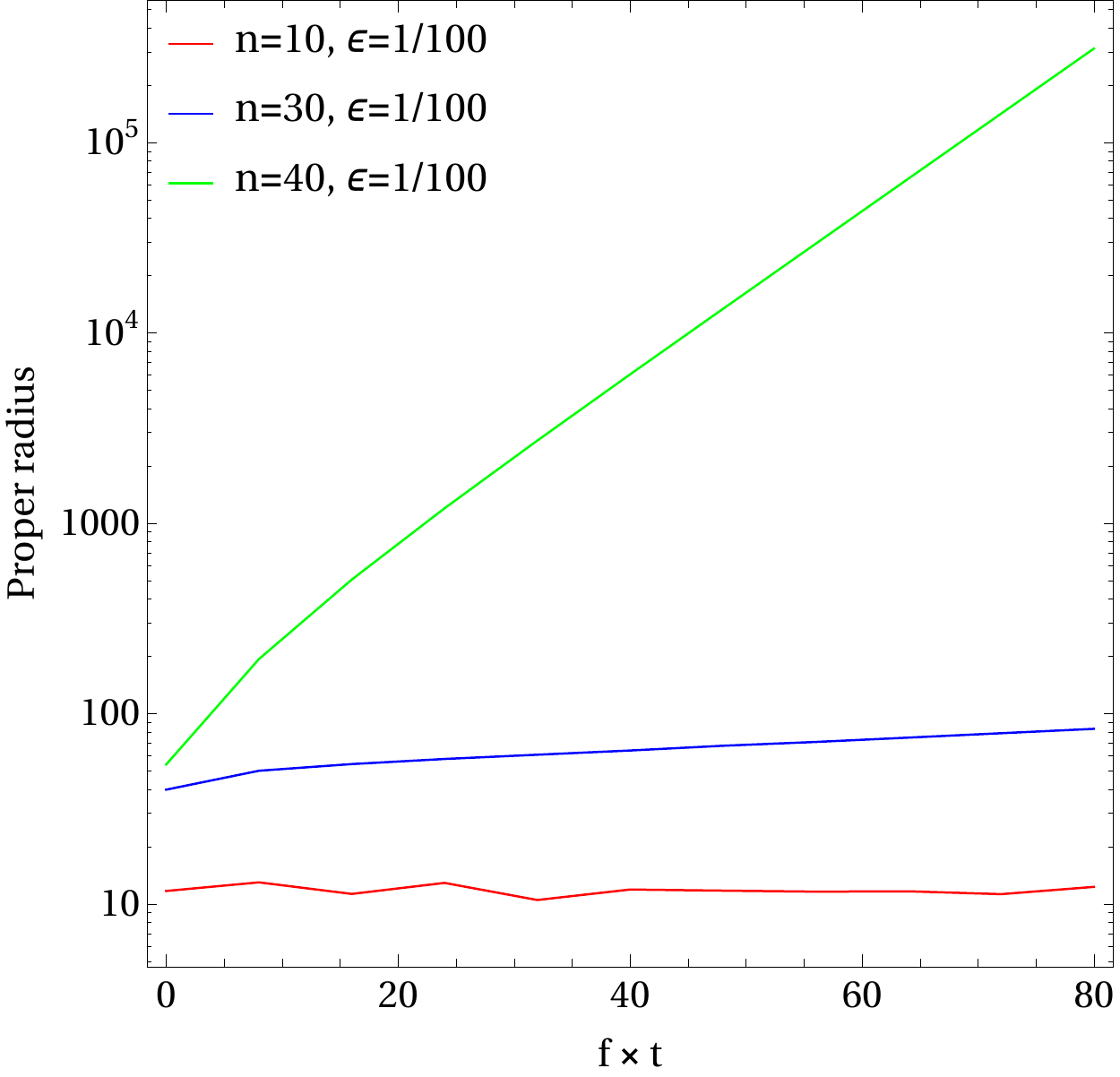}~~~~~
\includegraphics[width=0.45\linewidth]{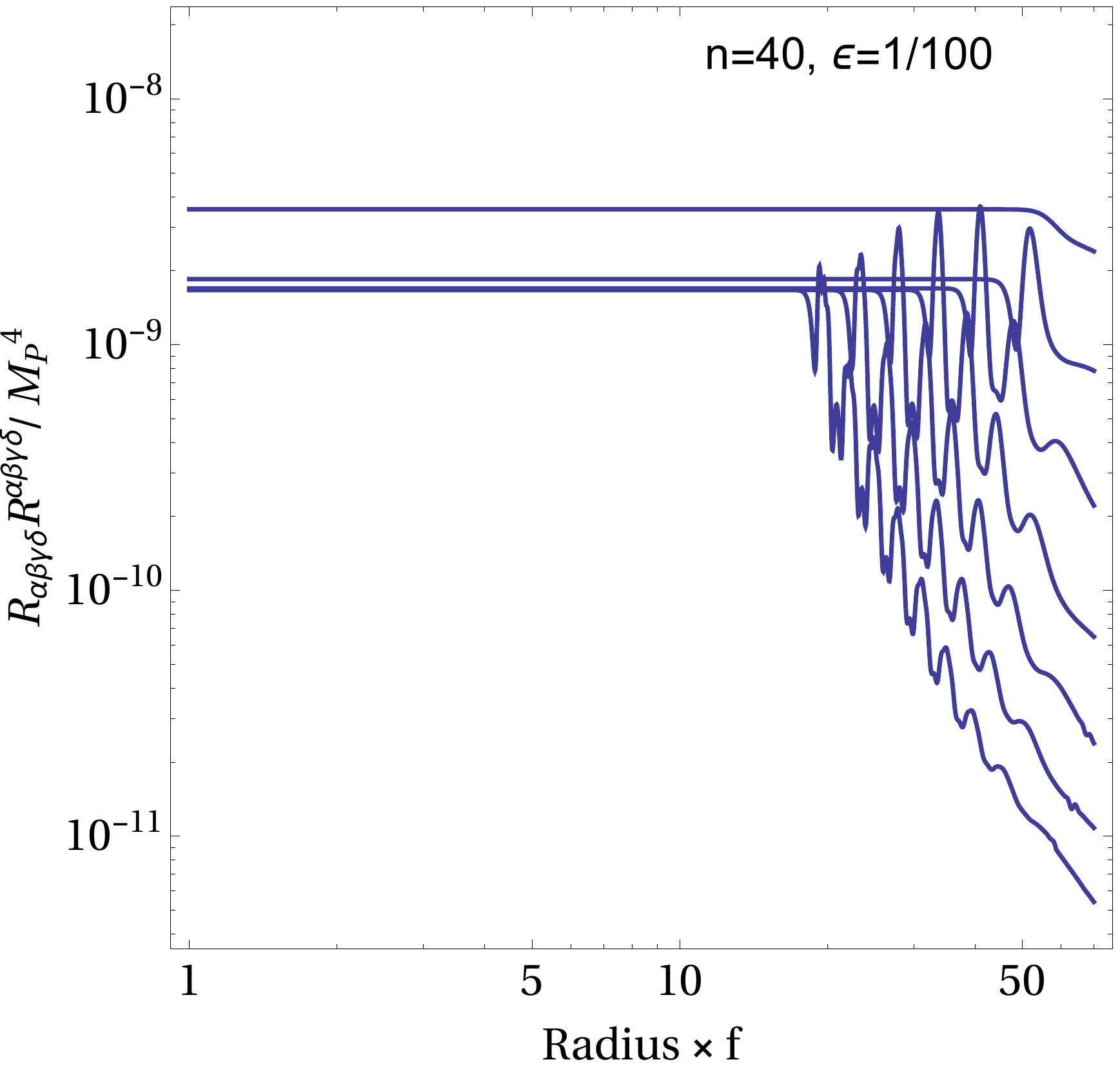}~~~~
\caption{Left: The proper radius of the locus $\phi=0.1f$ as a function of time for $n=10,30,40$ (from bottom to top) and $\epsilon=1/100$. For $n=40$ the core inflates. Right: The Kretschmann scalar curvature. Time starts at $0$ with the top curve, corresponding to a flat initial-value metric, and increases for each line below that, saturating at a substantially subplanckian value of order $\epsilon^4$ in the core.} 
\label{fig:inflatingcore}
\end{center}
\end{figure} 

\begin{figure}[t!]
\begin{center}
  \includegraphics[width=0.45\linewidth]{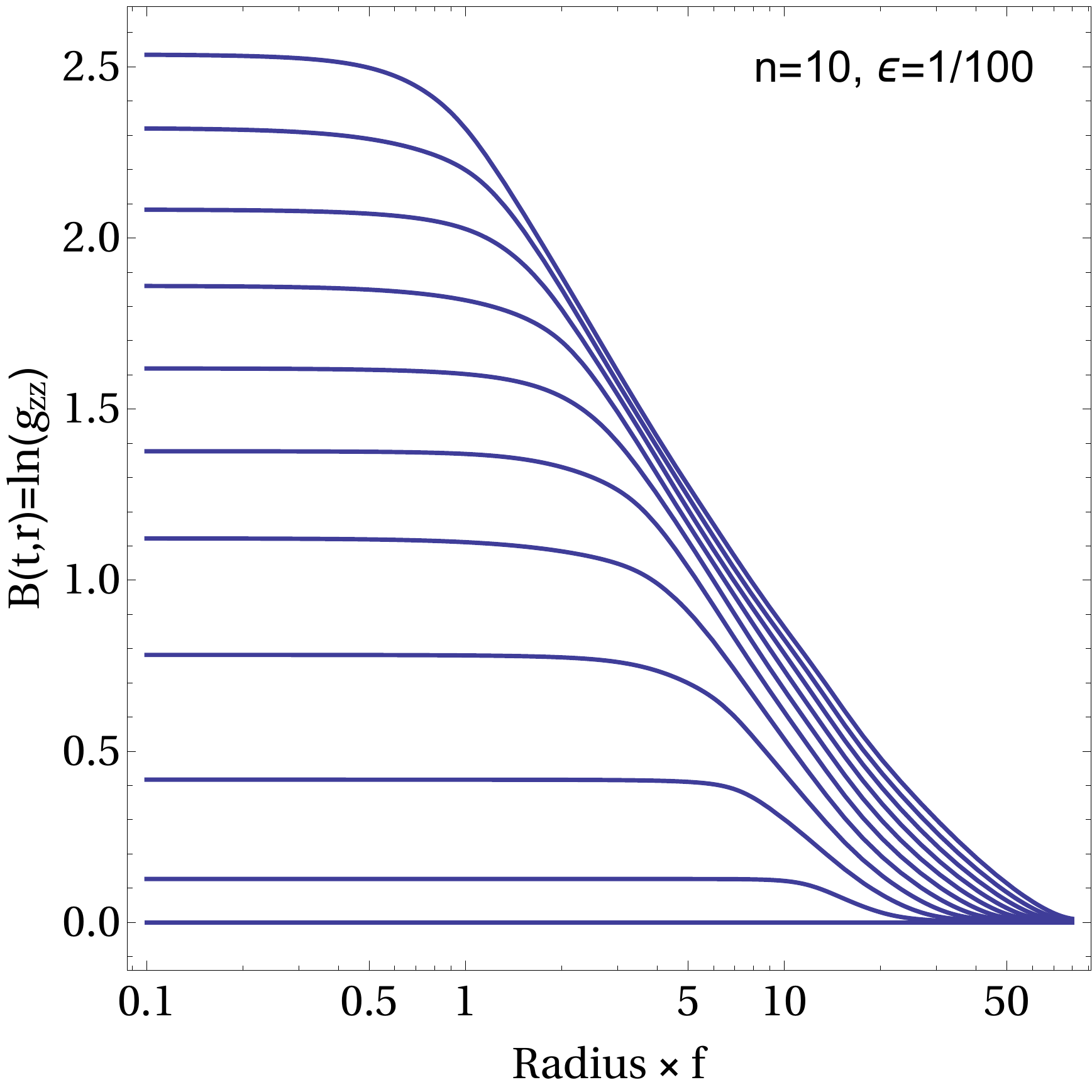}~~~~~
  \includegraphics[width=0.45\linewidth]{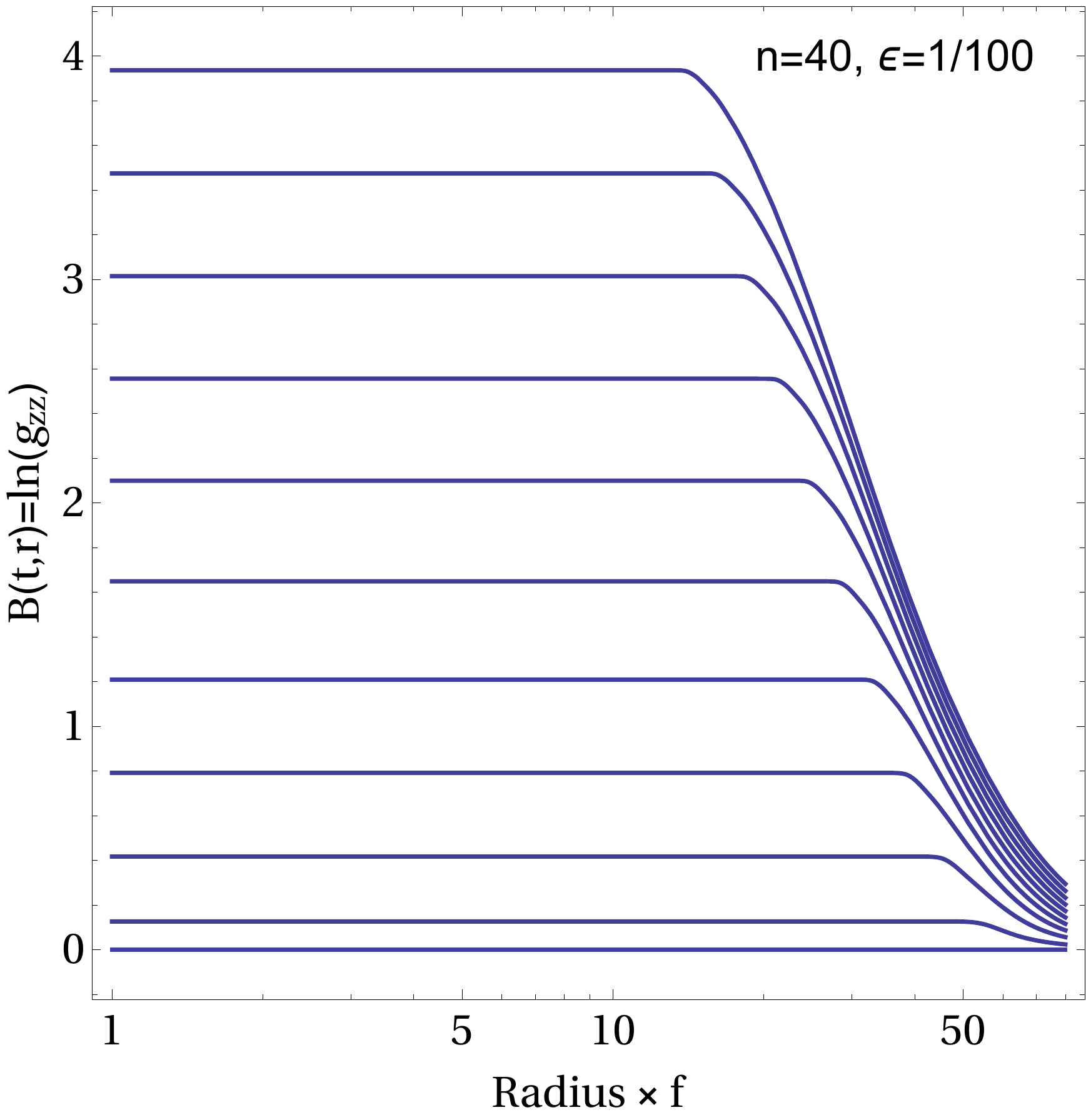}
  \caption{The evolution of the axial metric exponent $B(t,r)$  on a number of fixed time-slices
  for subcritical $n=10$ (left) and supercritical $n=40$ (right) strings with $\epsilon=1/100$. Time starts at $0$ with the bottom curve, corresponding to a flat initial-value metric, and increases for each line above that. In both scenarios, $g_{zz}$ grows exponentially with $t$, while the coordinate boundary of the core contracts exponentially with $t$. }
\label{fig:axialcomp}
\end{center}
\end{figure}

To check this picture of the cores of supercritical strings, we numerically solve the Einstein and scalar field equations in $\Phi^4$ theory with $n\gg 1$ and $\epsilon\ll 1$, starting from a flat space string profile and Minkowski initial condition. Our metric ansatz takes the form
\begin{align}
ds^2=dt^2-e^{H(t,r)}dr^2-e^{A(t,r)}r^2d\theta^2-e^{B(t,r)}dz^2\;.
\label{eq:metricansatz}
\end{align}
We show in Fig.~\ref{fig:inflatingcore} the proper radius of the locus $\phi=f/10$ as a function of time for $n=10,30$ and 40 and fixed $\epsilon =1/100$. We see that exponential expansion begins for sufficiently large $n$ (in this case between somewhere between $n$=30 and 40), while for lower values of $n$, the proper radius of the core is fixed. In Fig.~\ref{fig:inflatingcore} we also show the Kretschmann scalar to demonstrate that curvatures are small and the EFT is under control even in the topologically inflating regime.  Fig.~\ref{fig:axialcomp} shows the development of the logarithm of axial metric component $g_{zz}$ in both sub ($n=10$) and supercritical ($n=40$) scenarios on a number of fixed time-slices. Inside the core the $\log(g_{zz})$ is seen to grow linearly, corresponding to exponential growth. The collapse of the core suggested by the analysis above is also  visible in both cases.

\subsubsection{Wilson Loop Axions}

The analysis for $\Phi^4$ theory indicates that the onset of topological inflation is at least somewhat sensitive to the UV physics probed by the string core. We now turn to supercritical strings in Wilson loop axion models\footnote{For a more pedagogical discussion of this setup in the context of string theory, we refer the reader to Ref.~\cite{Silverstein:2013wua}.}.

The string is defined by the background value for the zero mode of $A^5$,
\begin{align}
A^5=n\theta/g\langle R\rangle,
\end{align}
where $g$ is the 4D $U(1)$ gauge coupling and R is the 5D radion/dilaton. The periodicity of the $A^5$ zero mode is $f\sim 1/gR$. In the 4D Einstein frame, the radion Lagrangian takes the form
\begin{align}
{\cal L}=\frac{3}{4} M_p^2\left(\frac{\partial R}{R}\right)^2-\frac{1}{4R}F_{\mu\nu}F^{\mu\nu}-V(R),
\end{align}
corresponding to a canonically normalized dilaton,
\begin{align}
R\rightarrow v_R e^{\sqrt{\frac{2}{3}}\frac{\phi}{M_p}},
\label{eq:dilaton}
\end{align}
where $v_R=\langle R\rangle$.

\begin{figure}[t!]
\begin{center}
\includegraphics[width=0.65\linewidth]{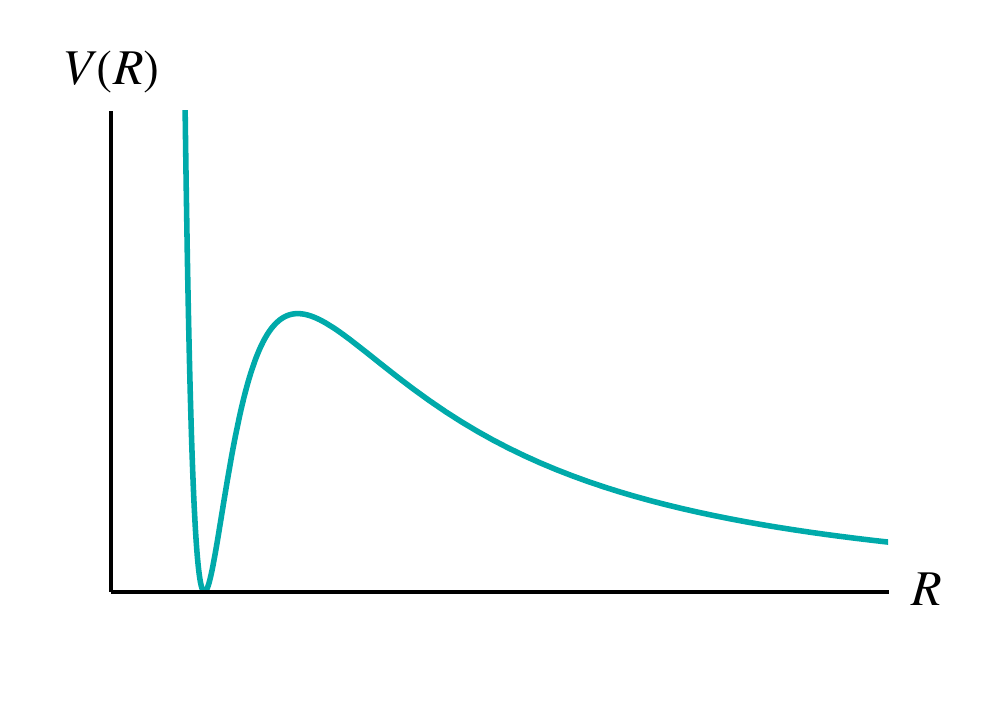}
\caption{Schematic form of a stabilized radion potential.} 
\label{fig:radpot}
\end{center}
\end{figure}

Common radion potentials arise from Casimir stabilization (see, e.g.,~\cite{ArkaniHamed:2007gg}) and the Goldberger-Wise mechanism~\cite{Goldberger:1999uk}. As $R\rightarrow\infty$, $V(R)$ exhibits qualitatively different behavior from the $\phi\rightarrow 0$ limit of $V(\phi)$ in $\Phi^4$ theory, falling to zero as an inverse power of $R$. A sketch of a radion potential is shown in Fig.~\ref{fig:radpot}. A typical way to obtain a Minkowski minimum at small $R$ is to tune a 5D cosmological constant, which then controls the asymptotics of the radion potential,
\begin{align}
V(R)\sim \frac{\Lambda v_R}{R}
\label{eq:Rasympt}
\end{align}
at large $R$, and $\Lambda$ is the c.c.

For simplicity, we will focus on $f<M_p$, where $M_p$ is the 4D Planck scale, but the case $R\ll M_p^{-1},~g\ll 1,~f>M_p$ is also interesting, corresponding to strings in extranatural inflation~\cite{extranatural}. The singular behavior of the axion gradient energy term near the core of a string is cut off by $R\rightarrow\infty$. It can be checked that $R$ diverges as $R\sim -n M_p \log(r)$ near the origin, rendering finite the core contribution to the total energy per unit length $\sim\int dr/r\log^2(r)$. Energy densities are still singular at $r=0$, and we imagine a UV cutoff somewhere between $R^{-1}$ and $M_p$.

In the flat-space limit, we might expect that the cosmic string interior would be separated from the exterior by a domain wall region corresponding to the peak in the potential separating the vacua at $v_R$ and infinity. From the outside, the domain wall begins at a radius determined by the balance of axion gradient and radion potential energies,
\begin{align}
r_{DW}\simeq \frac{n^2}{ g^2 M_p}
\end{align}
and has width determined by the balance of radion gradient and radion potential energies
\begin{align}
\sigma_{DW}\simeq v_R^2 M_p\;.
\label{eq:sigdw}
\end{align}
In these estimates we have approximated the potential in the peak region by $V\sim v_R^{-4}$.  The absence of a false vacuum at the core of the string leads to $r_{DW}\sim n^2$ instead of $n$ as in $\Phi^4$ theory.  This domain wall picture applies in the ``thin-wall" limit, corresponding to
\begin{align}
\frac{n}{g \cdot v_R}> M_p\;.
\label{eq:radioncondition}
\end{align}
This condition is analogous to the $n\cdot f> M_p$ condition for topological inflation in the case of $\Phi^4$ strings.
For smaller $n/gv_R$, the radion gradient term becomes more important than the potential, and the domain wall disappears. In this case the interior of the string is everywhere well-approximated by $R\sim -n M_p \log(r)$.

For large axion field excursions, we are interested in precisely the case~(\ref{eq:radioncondition}). What happens when we turn on 4D gravity? Although experience with $\Phi^4$ theory suggests that the spacetime of the supercritical string will be highly curved and nonstatic, the static flat-space structure described above is still useful for intuition. On one hand, due to the absence of a false vacuum in the string core, the arguments used to support topological inflation in the $\Phi^4$ string do not apply.  On the other hand, domain walls themselves are capable of realizing topological inflation, as in~\cite{Vilenkin:1994pv,Linde:1994hy}: might this occur for the radion string? This appears unlikely from Eq.~(\ref{eq:sigdw}), which suggests that the domain wall size does not grow with $n$. Instead, we expect the domain wall to have fixed proper width.

\begin{figure}[t!]
\begin{center}
\includegraphics[width=0.45\linewidth]{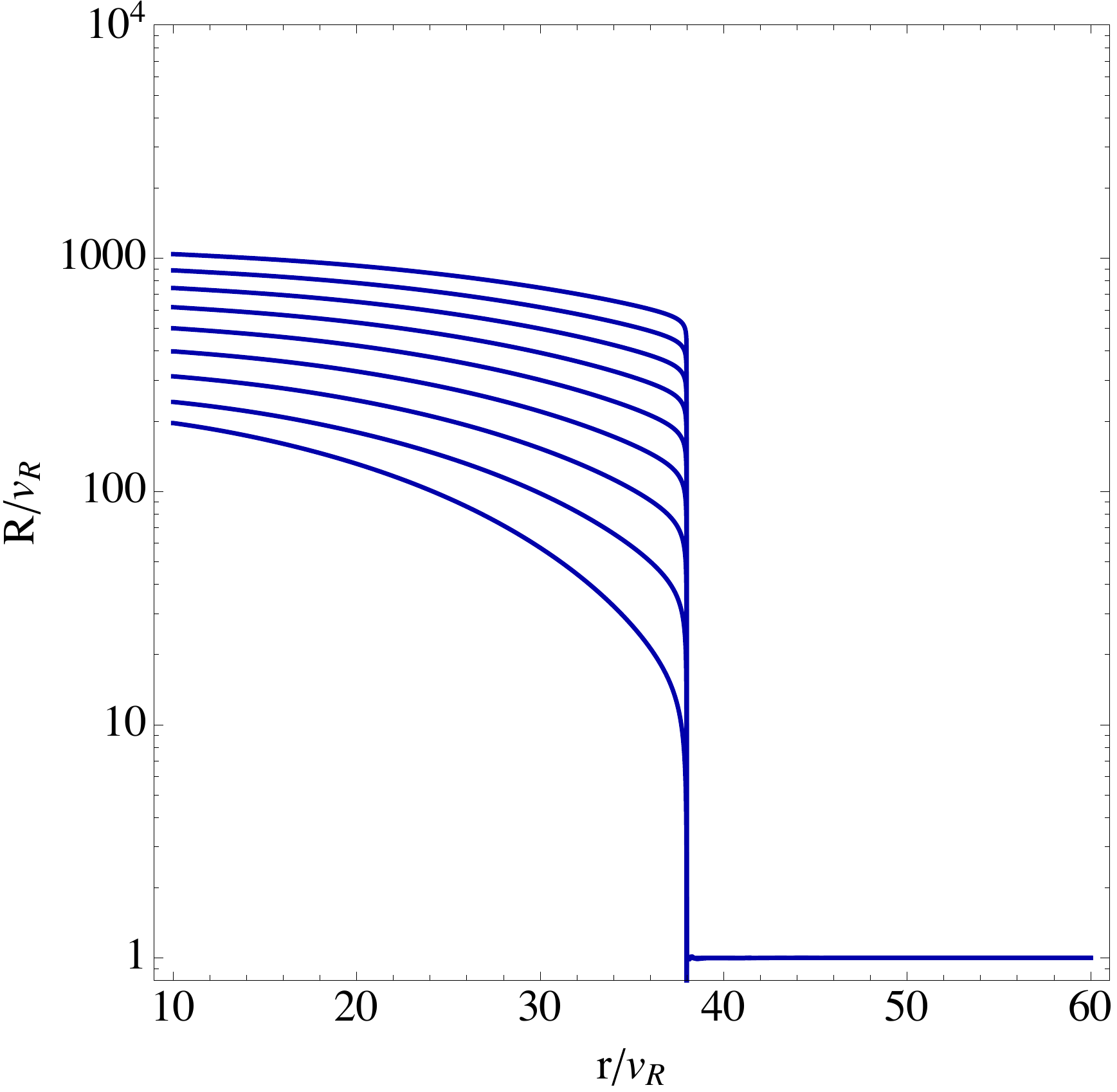}~~~~~
\includegraphics[width=0.45\linewidth]{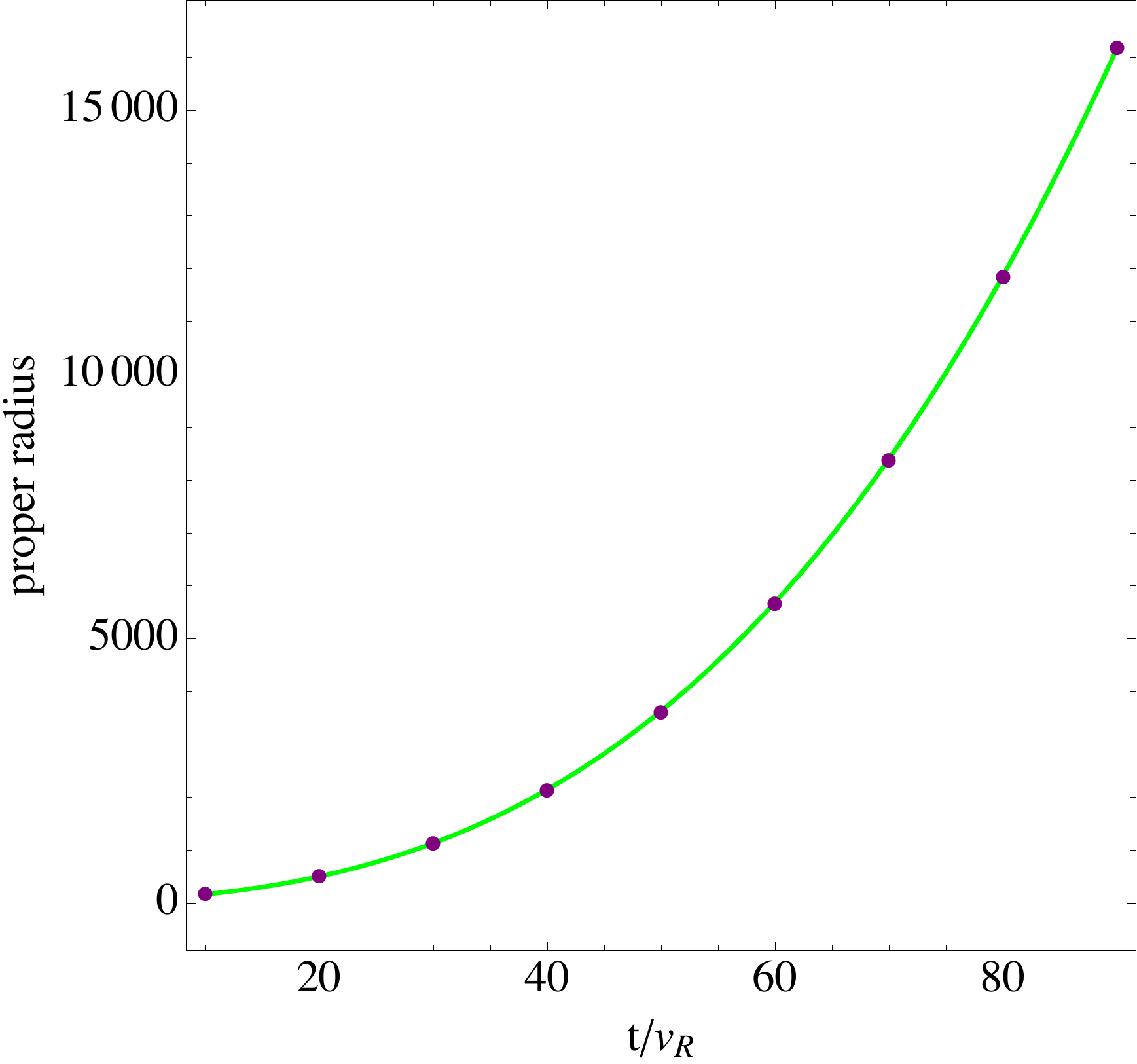}
\caption{Numerical analysis of a cosmic string in a sample 5D model with $g=1,~n=16,~v_r^2M_p^2=1/2,$ and radion potential $V=\frac{2}{R^6}-\frac{5}{R^3}+\frac{3}{R}$, in units where the Minkowski vacuum is at $v_R=1$. Left: Equally-spaced timeslices of the radial profile of the radion field (time runs bottom to top). A domain wall forms around $r\simeq 38$, outside of which $R$ takes its vacuum value. Right: Time dependence of the proper distance between $r=0$ and $r=37$ (purple dots), with a fit to a model of the form $c_1 (c_2+t)^3$ (green line).} 
\label{fig:radevol}
\end{center}
\end{figure} 

Let us consider the interior of the string more carefully, in a spirit similar to the previous section. If expansion occurs, the noncompact modulus becomes more homogeneous, and can be treated locally as a constant field rolling in its potential. In the string interior, which samples the asymptotic region of the potential, this corresponds to a dilaton~(\ref{eq:dilaton}) rolling in the potential~(\ref{eq:Rasympt}). Rolling dilaton cosmologies lead to power-law scale factors, and in the case studied here of a 5D dilaton in a c.c.-generated potential, we obtain
\begin{align}
a(t)\sim (t+c)^3\;,
\label{eq:at3}
\end{align}
where $c$ is a constant.
In this picture, the string core still inflates, but with {\emph{non-exponential}} accelerated expansion.

Does such expansion take place? To answer this question, we solve the Einstein-radion equations numerically as in the $\Phi^4$ case, taking sample values $g=1,~n=16,~v_r^2M_p^2=1/2$.  Fig.~\ref{fig:radevol} shows the radion evolution and the time dependence of the proper distance from the origin to the domain wall, comparing to a fit of the form~(\ref{eq:at3}). We find that the proper radius of the core indeed undergoes accelerated expansion, with Eq.~(\ref{eq:at3}) providing a good fit to the scale factor. We expect that this behavior persists to smaller $v_r^2M_p^2$, which correspond to models under better theoretical control, so long as Eq.~(\ref{eq:radioncondition}) is satisfied, but
we postpone a more comprehensive analysis of cosmic strings in models of this type for future work.

\subsection{Infinite String Exterior}
\label{sec:exterior}

We now turn our attention to the metric outside a topologically inflating core. We restrict our attention to the exponentially inflating case of $\Phi^4$ theory, deferring an analysis of the exterior of holonomy axion strings to future work.  

\begin{figure}[t!]
\begin{center}
\includegraphics[width=0.55\linewidth]{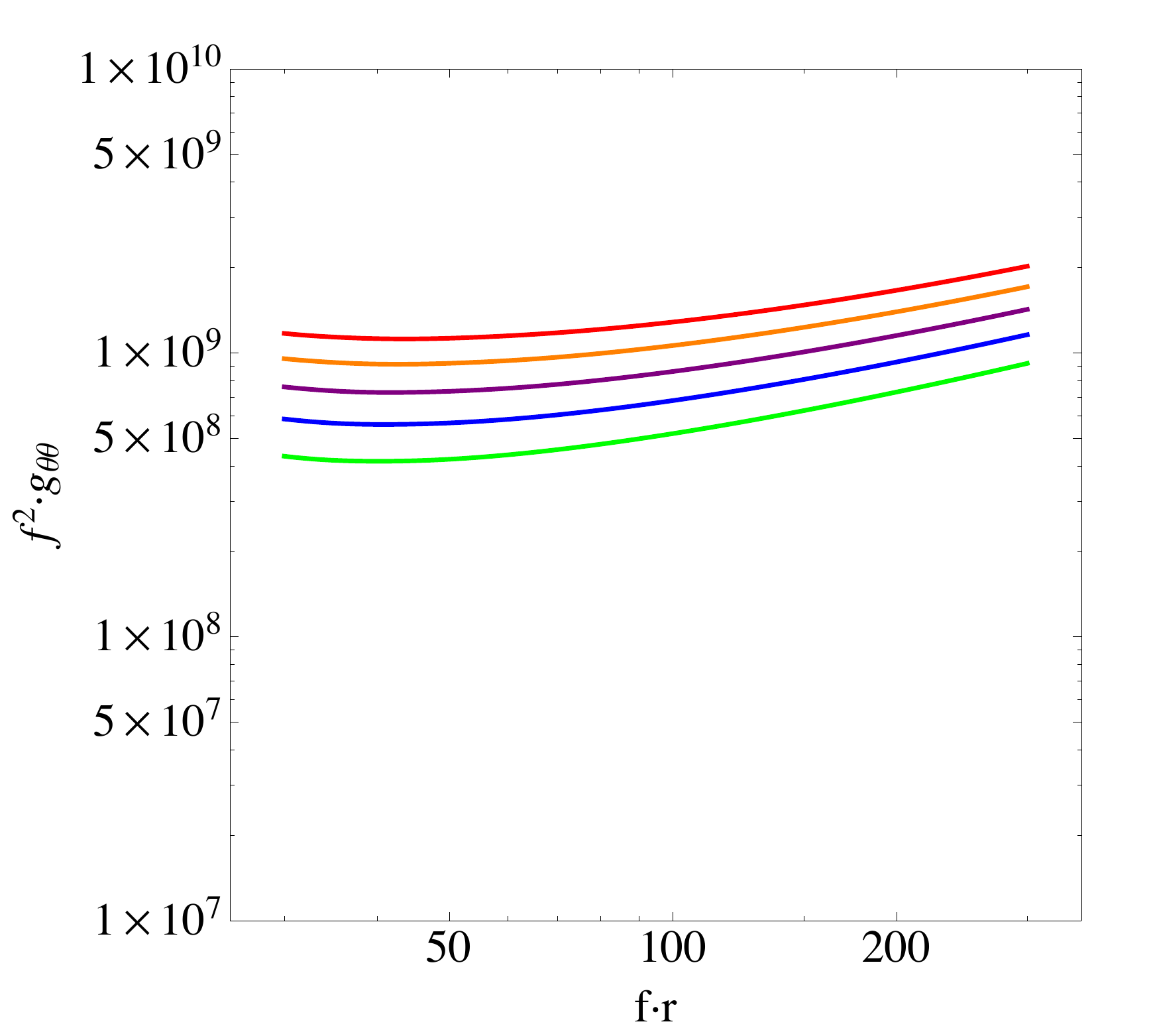}
\caption{Late-time behavior of $g_{\theta\theta}(t,r)$ exterior to the string core in $\Phi^4$ theory with $n=20,~\epsilon=1/10$. Lines represent equally-spaced $t$-slices of the numerical solution, advancing bottom to top.} 
\label{fig:thetaraw}
\end{center}
\end{figure}

A clue to the nature of the exterior geometry is apparent in the late-time behavior of $g_{\theta\theta}(t,r)$, plotted in Fig.~\ref{fig:thetaraw}. We see that $g_{\theta\theta}$ is nearly flat in $r$ at late times, indicating a cylindrical structure, or a cigar geometry, including the core at small $r$. 

A cylindrical ansatz suggests a reinterpretation of the supercritical string. In addition to cosmic strings, a closely related ``physical consequence" of the discrete gauge symmetry associated with an axion is the possibility of wrapping compact dimensions with axion flux. Here we see that far from the string core, these phenomena are equivalent. 

Consequently, we can derive the rest of the asymptotic exterior metric as a radion cosmology, where the radion rolls in a potential generated by axion flux. To see this, we parametrize the axisymmetric 4D line element as
\begin{align} 
ds^2_{(4)} = \frac{1}{R^2(t)}ds_{(3)}^2- R^2(t) d \theta^2\;,
\label{eq:4d_radion}
\end{align}
where the three-dimensional line element is taken to be of the form
\begin{align}
ds^2_{(3)}= dt^2 - a(t)^2 \left( dr^2 + g(r)^2 dz^2 \right)  \;.
\end{align}
The dimensionally-reduced action is then in the Einstein frame,
\begin{align}
S_{(3)} = \int d^3 x \sqrt{\operatorname{det} g_{(3)}} \left(\frac{f^2}{2 \epsilon} \Ricci_{(3)} + \frac{f^2}{\epsilon R(t)^2} g_{(3)}^{a b} \nabla_a R \nabla_b R - \frac{n^2 f^2}{R(t)^4} \right)\;,
\label{eq:S3}
\end{align}
where we have coupled gravity to the model~(\ref{phifour}) with $\phi\rightarrow f$. The axion flux generates a potential $V(R)\sim R^{-4}$. The equations of motion are
\begin{align}
R \dot{R}+R^3 \frac{\ddot{a}}{a} - \frac{n^2 \epsilon}{R} =& 0\\
-2 \frac{\dot{a}}{a}\dot{R} + 2\frac{n^2 \epsilon}{R^3} + \frac{\dot{R}^2}{R} - \ddot{R}=&0 \\
-2\frac{n^2 \epsilon a^2}{R^4} + \dot{a}^2+ a \ddot{a}-\frac{g^{\prime \prime}}{g} = &0,
\end{align}
and a solution is
\begin{align}
R(t) = \sqrt{2 n \epsilon^{1/2} t},\qquad a(t) = 2 t,  \qquad g(r) = f e^{-\sqrt{2} r }.
\label{eq:radion_components}
\end{align}
Inserting Eq.~(\ref{eq:radion_components}) into Eq.~(\ref{eq:4d_radion}) and making the coordinate transformations $r \rightarrow \log(f r),~t\rightarrow n\sqrt{\epsilon}t^2/2$, we obtain:
\begin{align}
ds^2=dt^2-t^2\left(r^{-2} dr^2+n^2\epsilon d\theta^2 + (f r)^{-2\sqrt{2}}f^2 dz^2\right)
\label{eq:ourmetric}
\end{align}
 Eq.~(\ref{eq:ourmetric}) satisfies the 4D Einstein equations coupled to the model~(\ref{phifour}) with $\phi\rightarrow f$. It satisfies the $\phi$ equation of motion to order $1/t^2$, and thus may be considered as the asymptotic behavior of the spacetime exterior to the string.

\begin{figure}[t!]
\begin{center}
\includegraphics[width=0.45\linewidth]{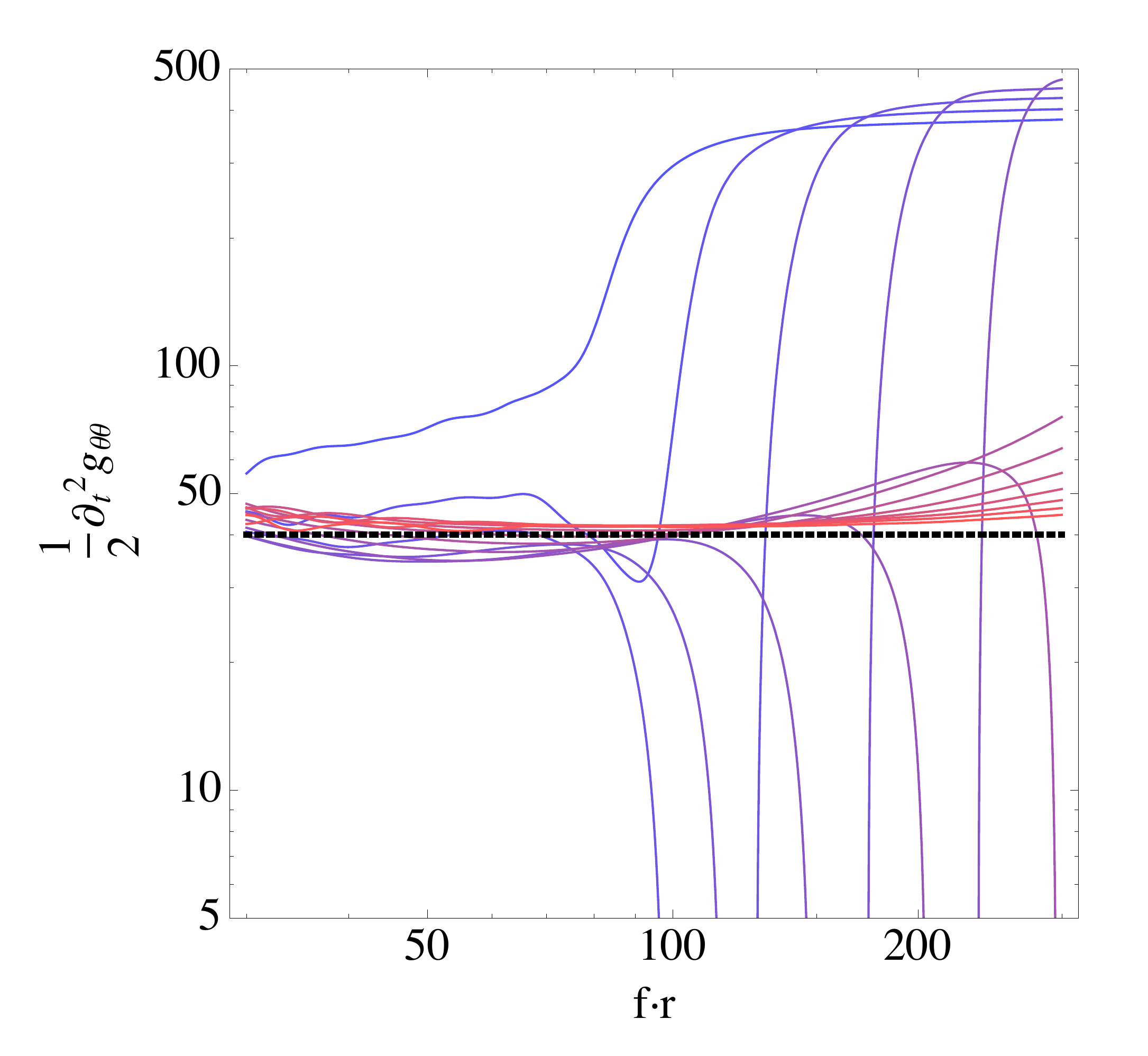}\qquad
\includegraphics[width=0.45\linewidth]{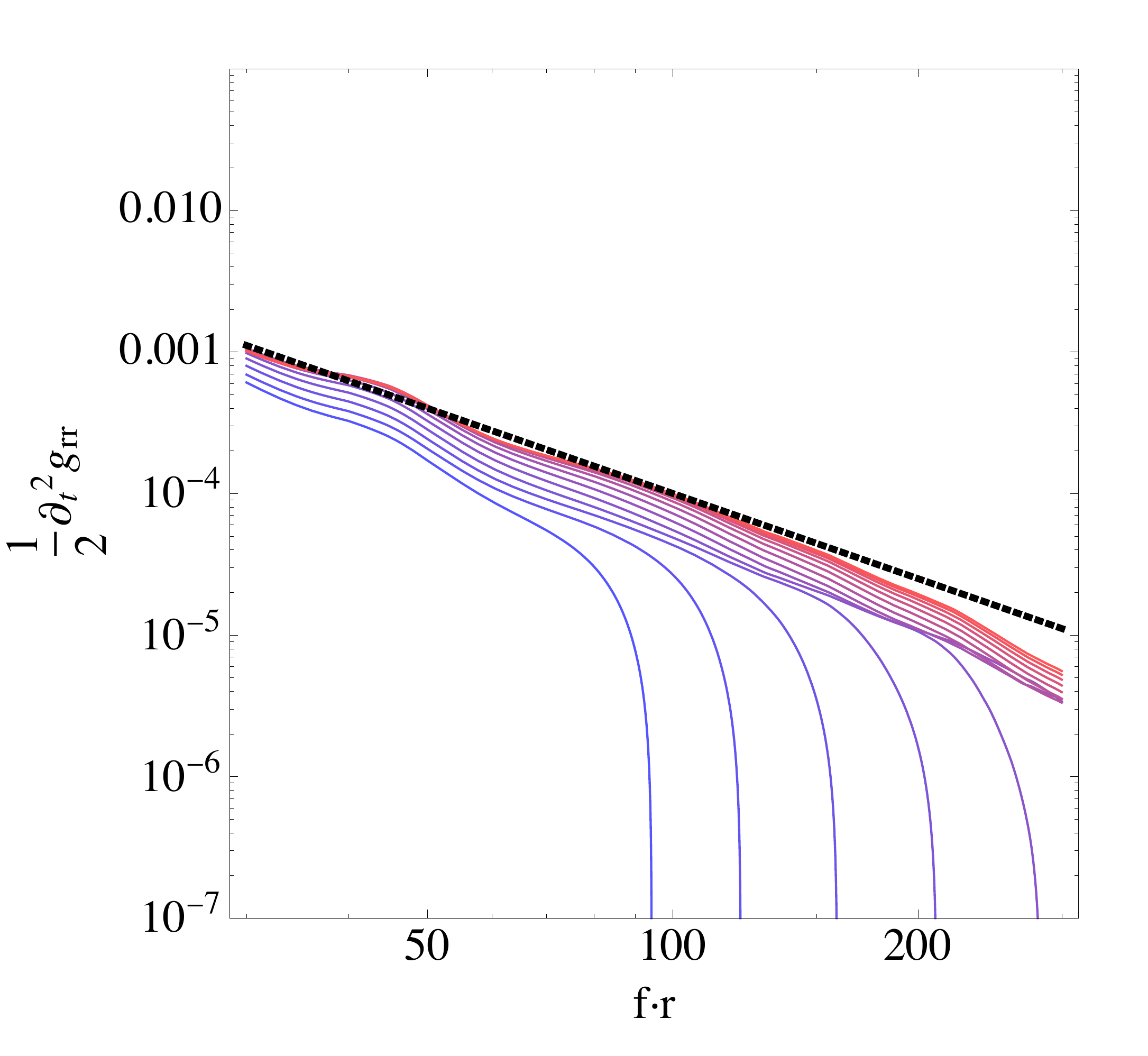}
\caption{Time evolution of $\partial_t^2 g_{\theta\theta}(t,r)$ (left) and $\partial_t^2 g_{rr}(t,r)$ (right) exterior to the string core with $n=20,~\epsilon=1/10$.  Solid lines are equally spaced $\log(t)$-slices of the numerical solution (early=blue, late=red).  At early times, a shock wave propagates outwards as the system adjusts away from the initial conditions.  At late times, the system converges toward the cylindrical metric given in Eq.~(\ref{eq:ourmetric}), indicated by the dashed lines.}
\label{fig:thetartt}
\end{center}
\end{figure} 

We can compare the metric~(\ref{eq:ourmetric}) to our numerical results with ansatz~(\ref{eq:metricansatz}) over a large range of $r$ exterior to the core region. We predict that the late-time behavior of each metric component scales as $t^2$, and that the constant value of $g_{\theta\theta}$ at fixed time is $n^2\epsilon$. In the left-hand panel of Fig.~\ref{fig:thetartt}, we show that both of these predictions are approximately satisfied by $g_{\theta\theta}$.  In the right-hand panel, we test the prediction $g_{rr}\sim t^2r^{-2}$. To match coordinate systems with the numerics, we transform $r\rightarrow r^\prime(r)$ so that $g_{zz}\sim r^{-2\sqrt{2}}$. We conclude that the cylindrical metric~(\ref{eq:ourmetric}) provides a qualitatively good fit to the late-time behavior, with some mild deviation at large $r$. The system appears to settle down to a cigar-like geometry from the flat-space initial conditions quite slowly, and we expect that simulations out to later times would improve the agreement with the metric in Eq.~(\ref{eq:ourmetric}).

\section{Measurability of the Transplanckian Excursion}
\label{sec:measure}
We now turn to the causal structure associated with supercritical strings, and ask whether an observer can access the large excursion of the axion field. Two simple thought experiments are:
\begin{itemize}
\item In the background of an infinite supercritical string, are there causal trajectories that circumnavigate the string?
\item In asymptotically flat space containing a large loop of string, are there causal trajectories that thread the loop and escape to infinity?
\end{itemize}
In either case, the trajectory would allow an observer to measure most or all of the transplanckian excursion of the axion.

With the exterior metric~(\ref{eq:ourmetric}), it is a simple matter to address the first question. Causal trajectories can indeed circumnavigate the string: for example, a family of null geodesics traces out circles of constant $(r,z)$. However, we note that from the point of view of stationary observers, it takes the signal an exponentially long time to complete the orbit.  Using $\Delta a = n f \Delta \theta$, we find that an observer that emits a signal at time $t_0$ receives it again at 
\begin{align}
t=t_0 e^{\frac{2\pi n f }{M_p}}\;.
\label{eq:exptime}
\end{align}
(Note that the big bang singularity of (\ref{eq:ourmetric}) must be cut off by the physical formation of the string, which prevents taking the $t_0\rightarrow 0$ limit in (\ref{eq:exptime}).) Although the time is finite, it is interesting that it is exponentially long precisely for supercritical strings. We have intentionally focused only on classical effects in this study, but on such timescales semiclassical nonperturbative phenomena may be important.

\begin{figure}[t!]
\begin{center}
\includegraphics[width=0.65\linewidth]{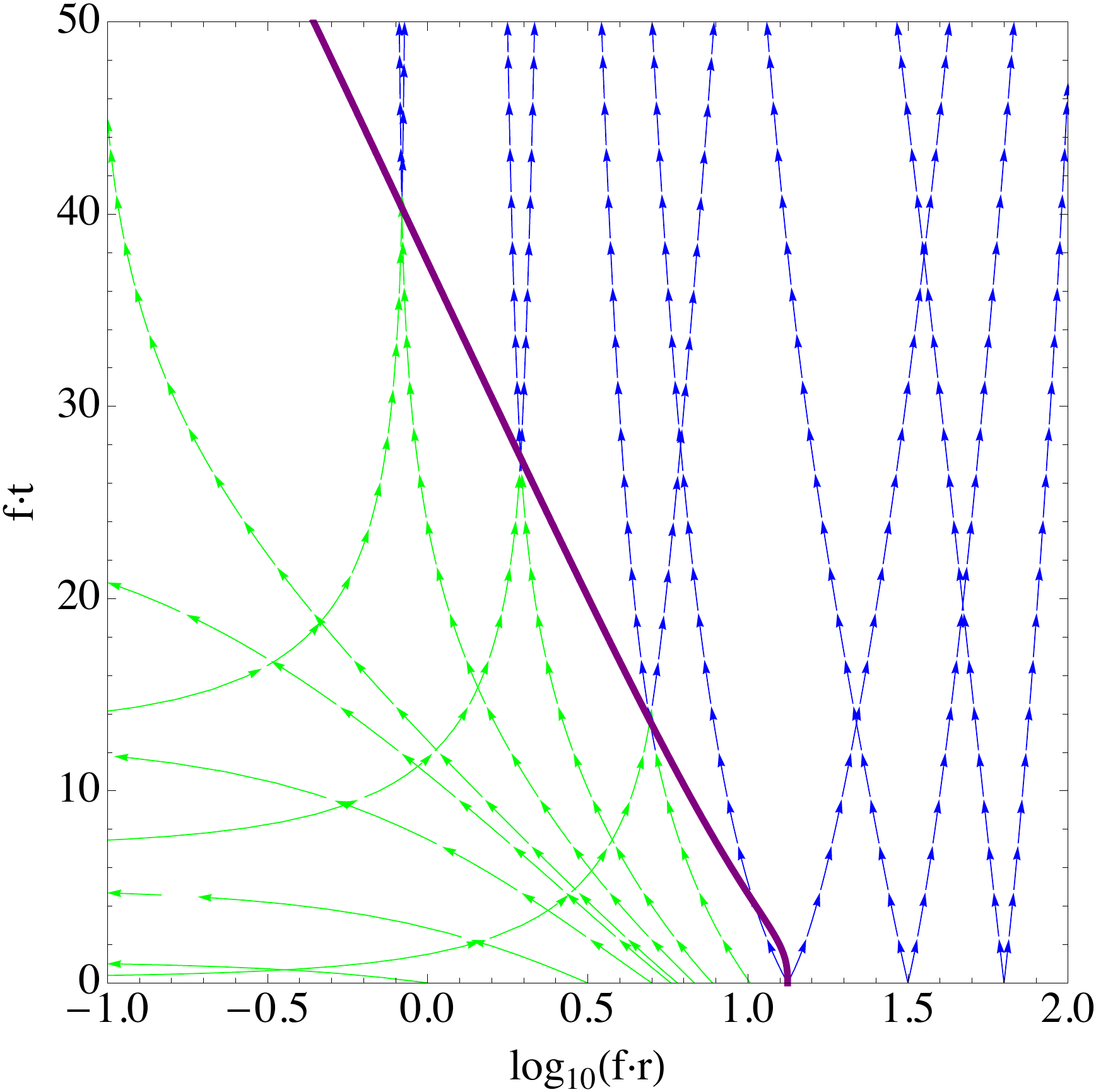}
\caption{Causal structure in $(r,t)$ for $n=8,~\epsilon=1/4$. The thick purple line tracks $X=f/2$ and represents a spacelike hypersurface. Arrowed contours trace future-directed null geodesics, which are colored green inside the core and blue outside it.}
\label{fig:causalstructure}
\end{center}
\end{figure} 

The second question, on the other hand, is associated with an interesting new obstacle. For definiteness, let us assume that the spacetime contains a loop of cosmic string, and that there is a causal trajectory $\gamma$ from past null infinity $\scriminus$ to future null infinity $\scriplus$ that threads through the string. An observer following $\gamma$ can measure most or all of the total axion field excursion around the core of the string. Let us further assume that when $n f>\mpl$, near the string the spacetime is locally described by the metric of the infinite straight string, with topological inflation in the core (either of exponential or power-law type). 

The worldvolume of the boundary of topologically inflating defects is expected to be a spacelike hypersurface in the causal past of exterior observers~\cite{Vilenkin:1994pv}. 
Points on the interior of the defect recede from an exterior comoving observer at speeds that quickly exceed the speed of light. (This conclusion holds both for exponential and power-law accelerated expansion of the core.) In Fig.~\ref{fig:causalstructure} we show a selection of null geodesics around the infinite string in the $\Phi^4$ model. The thick purple line tracks $X=f/2$ and is a representative spacelike hypersurface: we observe that null geodesics may exit the core, but not enter it.

Because the inflating region is separated from the exterior by  spacelike hypersurfaces in the causal past of exterior observers, $\gamma$ cannot be smoothly deformed into another causal trajectory that does not thread the string. Any family of curves realizing such a deformation would have to cross the de Sitter horizon of the inflating region.

However, the topological censorship theorem~\cite{Friedman:1993ty} states that (under some reasonable assumptions including global hyperbolicity, asymptotic flatness, and the averaged null energy condition)  every causal trajectory from $\scriminus$ to $\scriplus$ is homotopic to a trivial curve from $\scriminus$ to $\scriplus$. Thus, $\gamma$ is either prohibited by the topological censorship theorem, or the true spacetime violates the assumptions of the theorem.\footnote{The NEC should be automatically satisfied in classical spacetimes with scalar fields, and we have checked that it is satisfied in our numerical solutions.}

We cannot draw a firm conclusion about the fate of loops of supercritical string without knowledge of the dynamical evolution, which may require numerical simulation. We do not attempt such a simulation here. However, we note that in the case of toroidal black holes (horizons that have toroidal structure at early times), conflict with topological censorship is avoided by rapid contraction of the hole~\cite{Jacobson:1994hs,Shapiro:1995rr}. No causal observer is able to thread the torus before it collapses.

\section{Conclusions}
\label{sec:concl}
Global cosmic strings provide -- at least at the level of thought experiments -- an observable manifestation of the discrete gauge symmetries associated with axions. In large-field axion inflation models, supercritical strings are natural probes of the transplanckian field excursion that occurs during inflation. In simple EFTs realizing large-field inflation via axion monodromy, the relevant strings have large winding number and subplanckian symmetry breaking scale such that the axion field excursion is of order $n\cdot f\gtrsim M_p$. We have studied strings of this type and their classical spacetimes in two toy models, a 4D $\Phi^4$ theory and a 5D Wilson loop axion model. We find that supercritical string cores exhibit topological inflation for sufficiently large winding number, and the accelerated expansion is of exponential type in the $\Phi^4$ model and power-law type in the Wilson loop model. We also find a nonsingular candidate exterior geometry and show that it arises in the numerical solution of the $\Phi^4$ model. The exterior geometry is a cigar, making the winding of the axion around a string physically equivalent to the winding of an axion around a circle-compactified geometry sufficiently far from the core. 

We have also studied the causal structure around cosmic strings and the accessibility of the transplanckian excursion to exterior observers. 
We have found that infinite strings may be circumnavigated, but signals take an exponentially long time to complete an orbit. We have also argued that topological censorship implies that loops of string in asymptotically flat space cannot be threaded by causal observers if their cores topologically inflate.

It would be interesting to extend these results to include more detailed analysis of supercritical strings associated with Wilson loop axions, which we have only started to analyze in this work. In particular, we have not performed large enough simulations to determine whether the exterior geometry is also of cigar-type when the core undergoes power-law inflation. It would also be of interest to numerically investigate the causal structure and evolution of supercritical string loops. Finally, it would be useful to study quantum effects in the infinite supercritical string background, to see whether nonperturbative processes alter the long-time behavior.

\vskip 1cm
\noindent {\emph{Note added:}} We would also like to mention the recent work~\cite{Hebecker:2017wsu}, in which the relevance of global cosmic strings to the magnetic WGC is explored. The work of~\cite{Hebecker:2017wsu} is complementary to the studies performed here.

\vskip 1cm
\noindent
{\bf Acknowledgements:} 
MJD is supported by the Australian Research Council.
PD thanks Lorenzo Sorbo, Jennie Traschen, David Kastor, Szilard Farkas, Matt Reece, Alex Maloney, Scott Thomas, Michael Dine, and Nima Arkani-Hamed for valuable discussions.

\vskip 1cm

\appendix
\section{Large $N$ QCD(adj) Models}
\label{appx}
In this appendix we expand on a class of toy monodromy models closely related to the SQCD model of~\cite{dinedrapermonteux}.
We consider $SU(N)$ gauge theory with scale $\Lambda$,  $N_A$ massless adjoint Majorana flavors $\psi_i$, and an axion $a_0$ with decay constant $\Lambda<f_0<M_p$ coupling as
\begin{align}
{\cal L}\supset -\frac{c_0}{2}\cdot\frac{\alpha}{4\pi}\frac{a_0}{f_0}\FFd\;.
\label{eq:micro}
\end{align}
$a_0\rightarrow a_0+2\pi f_0$ is taken to be the discrete gauge symmetry associated with the axion, so $c_0$ is an integer.\footnote{$c_0=-2T(R)n_fQ_\lambda$ in the case where the axion is UV-completed by a scalar $\phi\rightarrow \frac{f_0}{\sqrt{2}}e^{\frac{ia_0}{f_0}}$ with Yukawa couplings to $n_f$ Weyl fermions $\lambda_i$ in representation $R$ of the gauge group and carrying charge $Q_\lambda=-Q_\phi/2$ under the anomalous PQ symmetry. If $Q_\phi$ is normalized to $1$, the discrete gauge symmetry is normalized to $a_0\rightarrow a_0+2\pi f_0$. The fact that $c_0$ would be fractional for odd numbers of $SU(2)$ Weyl fundamentals, explicitly breaking the discrete gauge symmetry of the axion, is a manifestation of the Witten anomaly. $c_0=1$ for a single Dirac fermion of charge $-1/2$.} 

This theory has an $SU(N_A)\times U(1)_X\times {\mathbb{Z}}_{N}\times {\mathbb{Z}}_{c_0}$ global symmetry, where the Abelian part acts as
\begin{align}
U(1)_X&:~~a_0\rightarrow a_0+\beta\cdot \frac{NN_A}{d} f_0\;,\nonumber\\
           &~~~~\psi_i\rightarrow e^{i\beta \frac{c_0}{2d}}\psi_i~~~~~~~~~~~~~~~~~~~\beta\in[0,2\pi)\;\nonumber\\
{\mathbb{Z}}_{N}&:~~\psi_i\rightarrow e^{i \pi\frac{ q}{N}}\psi_i~~~~~~~~~~~~~~~~~~~q=0\dots N-1\;\nonumber\\
{\mathbb{Z}}_{c_0}&:~~a_0\rightarrow a_0+2\pi\frac{ p}{c_0}\cdot f_0~~~~~~~~~~p=0\dots c_0-1\;.
\end{align}
$d$ denotes the greatest common divisor of $NN_A$ and $c_0$, and the $U(1)_X$ charges have been normalized so that $2\pi$ is the smallest value of $\beta$ for which the transformation is a discrete gauge symmetry. A simple case is obtained by taking $c_0=d=1$.

Below $\Lambda$, strong dynamics is believed to lead to fermion condensation,
\begin{align}
\langle\psi_i\psi_j\rangle\sim N^2\Lambda^3 e^{i\frac{2\pi k}{N}}\delta_{ij}\;,
\label{eq:condensate}
\end{align}
up to a $U(1)_X$ transformation. The ${\mathbb{Z}}_{N}$ is spontaneously broken by the condensate and  $k=0\dots N-1$ labels the vacua. 
The coefficient in Eq.~(\ref{eq:condensate}) indicates the large $N$ scaling for a (canonically normalized) adjoint condensate. 

The low energy theory contains $\frac{1}{2}(N_A^2+N_A-2)$ massless pions, reflecting the $SU(N_A)/SO(N_A)$ pattern of chiral symmetry breaking, and the decay constants scale as $f_\pi\sim N$ in large $N$.   There is also a $U(1)_X$ axion $a$, satisfying
\begin{align}
\langle 0| J_X^\mu|a\rangle &= f p^\mu\;,\nonumber\\
J_X^\mu=\frac{N N_a}{d} f_0 \partial^\mu &a_0+\frac{c_0}{2d} \psi^*_i i\sigma^\mu \psi_i\;.
\end{align}
$a\rightarrow a+2\pi f$ is a discrete gauge symmetry in the infrared theory. 
 For $f_0\gg\Lambda$, there is little mixing between the states annihilated by the fermion and boson contributions to the current, and the light degree of freedom is mostly $a_0$. Therefore,
\begin{align} 
f= \frac{N N_a}{d} f_0+{\cal O}\left(\frac{\Lambda}{f}\right)\;.
\label{eq:feff}
\end{align}
From the form of the $U(1)_X$ transformation, we see that adjacent condensate vacua~(\ref{eq:condensate}) are smoothly connected under the motion $a\rightarrow a+2\pi f \frac{d}{N c_0}$. Each $\psi\psi$ branch is traversed $c_0/d$ times over the field range of $a$. From the point of view of the microscopic axion $a_0$, the monodromy group is $\mathbb{Z}_{NN_A/d}$. Thus in terms of microscopic angle-valued fields $a_0/f_0$ and ${\rm arg}(\psi\psi)$, the low-energy axion moduli space is a $(p,q)$ torus knot, where
\begin{align}
(p,q)=\left(\frac{NN_A}{d},\frac{c_0}{d}\right)\;.
\end{align}

The point of this construction is Eq.~(\ref{eq:feff}), which shows that for fixed microscopic scales $f_0$ and $\Lambda$, the field range of $a$ can grow with $N$. It is necessary that $c_0$, and thus $d$, is not proportional to $N$. (Although $f$ also grows with $N_A$, $N_A$ is bounded from above by an $N$-independent constant determined by either the onset of the conformal window or the loss of asymptotic freedom.)

More generally, parametrically large axion monodromy is produced by the presence of two global nonlinearly realized $U(1)$ symmetries with anomaly coefficients of different orders in $1/N$. The considerations above essentially restrict the $U(1)_{a_0}$ anomaly to be of order $g^2\sim1/N$ and the axial $U(1)_\psi$ anomaly to be of order $g^2N\sim1$. Thus the acceptable $\psi$ representations include the adjoint and the two-index tensors.

The situation described here is markedly different from what happens in large $N$ QCD with massless fundamental flavors and an axion. In that case there is a light $\etap$ boson that mixes with the axion, but no parametrically large monodromy in the axion direction. (There can be monodromy if the flavors are more massive than the anomaly -- in this case we return to the pure QCD description, with branches controlled by $\etap$ vevs.)

The simplest way to introduce a slowly varying potential $V(a)$ is to explicitly break the ${\mathbb Z}_N$ symmetry with small fermion mass terms,
\begin{align}
{\cal L}\supset -\frac{1}{2}(m_{ij}\psi_i\psi_j+cc)\;.
\end{align}
Then the potential term in the leading-order chiral lagrangian is
\begin{align}
V(\pi^a,a)=-\frac{1}{2}\mu \fpi^2 \left(\TR\left[mU\right]e^{ic_0a/f}+cc\right)\;,
\label{eq:chpt}
\end{align}
where $\mu\sim 4\pi \fpi/N$ and $U=e^{2i\pi^a T^a/f_\pi}$, $T^a\in su(N_A)/so(N_A)$. The $\pi$ fields may be integrated out to obtain $V(a)$.
For example, for $N_A=2$, the axion potential takes the form of the ordinary QCD axion potential with two light fundamental flavors:
\begin{align}
V(a)\propto N^2\Lambda^3\sqrt{m_1^2+m_2^2+2m_1m_2\cos(c_0a/f)}\;.
\end{align}

This is no longer exactly the natural inflation potential, and we could ask the conditions under which the model inflates. For $m_1\ll m_2$ the conditions on $\epsilon,f$ reduce to those of the 1-flavor model, $f/c_0 \gg M_p$. For $m_1=m_2$, there are quantitative but not qualitative differences.

\bibliography{cosmicstrings}
\bibliographystyle{JHEP}

\end{document}